\documentclass[twocolumn]{article}
\usepackage{amsmath}
\usepackage{amssymb}

\title{\textbf{Disentangling Nonlocality and Teleportation}}

\author{Lucien Hardy\thanks{\texttt{hardy@qubit.org}}\\
\textit{Centre for Quantum Computation,}\\
\textit{The Clarendon Laboratory,}\\
\textit{Parks road, Oxford OX1 3PU, UK}}

\begin{document}

\maketitle

\begin{abstract}
\noindent
Quantum entanglement can be used to demonstrate nonlocality
and to teleport a quantum state from one place to
another.  The fact that entanglement can be used to do both these things
has led people to believe that teleportation is a nonlocal effect.  In
this paper it is shown that teleportation is conceptually independent of
nonlocality.  This is done by constructing a toy local theory in which
cloning is not
possible (without a no-cloning theory teleportation makes limited sense)
but teleportation is. Teleportation in this local theory is
achieved in an analogous way to the way it is done with quantum theory.
This work provides some insight into what type of process teleportation is.
\end{abstract}

\section{Introduction}

Entangled states in quantum theory can be used both to demonstrate that
quantum theory is not a local hidden variable theory \cite{bella}
and to teleport
quantum states from one place to another \cite{teleport}.
The fact that entangled
states display nonlocality has led people to assume that teleportation,
which requires the use of entangled states,
is essentially a nonlocal phenomena.  This may be true of quantum
teleportation, but, as will be shown in this paper, it is not necessarily
true of teleportation in general.  To prove this a toy physical theory
will be described which has the following properties: (a) It is local,
(b) it does not allow cloning of states, and (c) it allows teleportation
of states.

The question of whether teleportation is essentially nonlocal has been
raised in two papers.  Popescu \cite{popescu} showed that certain mixed
states called
Werner states which have been shown by Werner not to violate any Bell
inequalities can nevertheless be used to teleport quantum states with a
greater fidelity than could be achieved without any entanglement.  More
recently, Braunstein \cite{braunstein} has shown how teleportation is
possible with
continuous variables.  He points out that it is possible to consider
only states having positive Wigner
functions and that the measurements he considers can all be functions of
$q$ and $p$.
If we are restricted to such measurements on a state with a positive
Wigner function then Bell \cite{bellb}has showed that it is possible to construct a
local hidden variable model.  However, it is possible to consider
different states and different measurements, and then there will be
nonlocality (see \cite{cohen}).

In general, it is very difficult to differentiate between two physical
concepts within the same physical theory.  If that physical theory has
both of the corresponding properties then these properties are likely to
be manifest at the same
time making it difficult to disentangle them.  However, by considering
alternative physical theories, we can hope to
show that two physical concepts are distinct

It is the no-cloning theorem which makes quantum teleportation
interesting.  In standard classical physics it is possible in principle
to make measurements on a system which establish a complete
description of the state without disturbing the system.  This complete
description of the state can then be used to build another system in the
same state.  Hence, a possible teleportation scenario would be that such
a complete measurement is made on a classical system, the resulting
information
is sent to another location, and a new copy is created.  The original
copy could be destroyed and this would seem like teleportation.
However, there is no reason to destroy the original. On the other hand,
if there is a no-cloning theorem which says it is impossible to make a
copy of a system when the state of the system is unknown then there can be
no dilemma about whether to destroy the original. It must have been
destroyed if there is now a copy somewhere else.  Hence we will take the
existence of a no-cloning theorem to be part and parcel of what we mean
by teleportation.  Since classical physics as it stands does not have a
no-cloning theorem, we will invent an alternative physical theory which
does and which is explicitly local. This theory is only intended to
illustrate the central point of this paper (it does not correspond to
any real physical system).

A number of real optical experiments have been performed to demonstrate
quantum teleportation \cite{expmts}.  Since these are real experiments the
detectors are less than 100\% efficient. Taking advantage of this fact
Risco-Delgado \cite{ramon} has shown that it is possible to build a local
hidden variable model which agrees with the results the first two of
these experiments (he has not considered the third).  However, this
model does not support perfect teleportation and it
is not it clear whether a no-cloning theorem can be derived. Hence,
they are not useful for our present purpose.
Nevertheless, they played a role in motivating the approach taken
here.

\section{A toy local theory}

We will now state the  seven postulates of the toy local theory.

\subsection{State description postulate}

\begin{description}
  \item[Postulate 1]  A system consists of particles and each
particle can be in one of four states labeled $0,1,2,3$.
\end{description}

A system of $N$ particles has the state $X=(x_1,x_2,\dots,x_n,\dots,x_N)$
where $x_n\in\{0,1,2,3\}$ is the state of the $n$th particle.  The
number of possible states $X$ is $4^N$.

\subsection{State measurement and preparation postulates}

\begin{description}
  \item[Postulate 2] There exist measurement apparatusses which can be used to
extract information about the state of a system (though, due to
postulates 4 and 5 below, only partial information can be extracted).  A
given measurement, $A$, will have $R$
outcomes labeled $r=0,1,2,\dots,R-1$.  Associated with the $r$th outcome is
the set of possible states
\[ A_r=\{ X_{r1},X_{r2},\dots,X_{rL} \}  \]
A given $X$ does not appear in more than one such set.  I.e. the sets
$A_r$ are disjoint.  Furthermore, each possible $X$ must appear in one of the
sets $A_r$.
  \item[Postulate 3] If the state of the system is $X$ then the outcome of a
measurement  of $A$ will be $r$ where $X \in A_r$. Since $X$ only
appears in one
of the sets $A_r$ there can only be one outcome for a given $X$.
  \item[Postulate 4] After a measurement of $A$ having outcome $r$ the
state of the
system will be $X_r$ with probability ${1\over L}$ (i.e. the state is
selected randomly from the set $A_r$).
\end{description}
>From postulates 3 and 4 we see that if a measurement is repeated on a
system the outcome will be the same.
Postulate 4 has the consequence that a measurement will, in general, disturb the state
of a system.  This limits the amount of information
that can be extracted about the state of a system by repeated
measurements.  The set $A_r$ can be represented by the matrix
\[ A_r=
   \left(
          \begin{array}{cccc}
                             x_1^1 & x_2^1 & \cdots & x_N^1 \\
                             x_1^2 & x_2^2 & \cdots & x_N^2 \\
                             \vdots & \vdots &\ddots & \vdots \\
                             x_1^L & x_2^L & \cdots & x_N^L
          \end{array}
   \right) \]
The entries in the $l$th row are the elements of $X_{rl}$ and the $n$th
column pertains to the $n$th particle.  For $L=5$ a typical column might be
\begin{equation}\label{columna}
\begin{array}{l} 2 \\ 0 \\ 0\\ 2 \\ 2 \end{array}
\end{equation}
\begin{description}
\item[Postulate 5]  The sets $A_r$ can be chosen in any way which is consistent
with the constraints imposed in the postulate 2 and the
following additional constraints.
The sets $A_r$ must be such that (i) in each column there are at
least two different values of $x$ and (ii) each value which occurs must occur
in the column for at least 25\% of the entries.
\end{description}
To illustrate postulate 5 we see that the column in (\ref{columna}) is ok
because (i) there are at least two different values, in this case 0 and 2,
and (ii) because each value which occurs  does occur for at least $1\over 4$
of the entries, in this case the ratios are $2\over5$ and $3\over5$.
Postulate 5, like postulate 4, limits the amount of information that can
be extracted by measurement.  If all the entries in a particular column
were 0 then, if that outcome were recorded, we could be certain that the
state of the corresponding particle is 0 both before and after the
measurement.  However, this is not possible.  Postulate 5 implies that
there must be at least a 25\% chance that the state is something else.

\subsection{State Manipulation postulates}

It is possible to manipulate the particles which make up a system though
only in accordance with the following rules.
\begin{description}
\item[Postulate 6] The particle can be moved around in space (though at
sub-luminal speeds).  A joint
measurement can only be made on two or more particles if they in the
same place.
\end{description}
This postulate is necessary since we are constructing a local theory.
\begin{description}
\item[Postulate 7] There exist apparatusses which make it possible to manipulate
each individual particle without knowing its state in such a way that if
the particle is in the state $x$ it will go to the state $U(x)$ where
$U$ is a one to one function.
\end{description}

These manipulations are taken to be analogous to unitary transformations
in quantum theory. We will be particularly interested in the operation
$U_k$ where
\begin{equation}\label{uk}
U_k(x)=(x+k)\mod 4
\end{equation}
We will call such operations {\it rotations}.

\section{Examples of measurements}

\subsection{One particle}

Consider making measurements on a single particle.  In this case $N=1$
so there will only be one column in the matrix representation of the
sets $A_r$.  One possible measurement is defined by the matrices
\[  A_0=\left( \begin{array}{c} 0 \\ 1 \end{array} \right)
    \qquad
    A_1=\left( \begin{array}{c} 2 \\ 3 \end{array} \right)  \]
We see that both $A_0$ and $A_1$ are consistent with postulates 2 and 5.
Imagine that Alice is given a single particle which has $x=0$ but that
she does not know what
its state is.  She could perform the measurement described by these
matrices.  Since $x=0$ she will get outcome 0 corresponding to $A_1$ by
postulate 3. She will now know that the original state was either 0 or 1
but she will not know which.  After this measurement the new state of
the particle could be either 0 or 1 (with even probabilities) by
postulate 4 but Alice will not know which.
She could repeat the measurement many times and each time she would get
the same outcome.  The state $x$, is in some
sense, a hidden variable since a particle cannot be prepared with a
known $x$.  The preparable states of the system are
combinations such as [50\% of 0 and 50\% of 1].  An alternative
measurement is defined by the matrices
\[  B_0=\left( \begin{array}{c} 1 \\ 2 \end{array} \right)
    \qquad
    B_1=\left( \begin{array}{c} 3 \\ 0 \end{array} \right)  \]
If Alice now gives the particle to Bob and he measures $B$ there is a
50\% chance he will get outcome 0 and a 50\% chance he will get outcome
1.  If he gets outcome 0 then since the state Alice gave him was
known to be [50\% of 0 and 50\% of 1] they can conclude that the state of
the particle Alice gave him {\it was} $x=1$.  However, the state of the
system after Bob's measurement will be [50\% of 1 and 50\% of 2]
corresponding to $B_0$.  Furthermore they do not know whether $x=1$ was
the state of the system that was originally given to Alice (in our
example it was not).

\subsection{Two particles}\label{bell}

The state of two particles is represented by $X=(x_1,x_2)$.  We will
be interested in only one possible measurement that can be made on two
particles.  This is described by the matrices
\begin{equation}\label{bstatea}
B_0=\left( \begin{array}{cc}  0&0\\ 1&1\\ 2&2 \\ 3&3\end{array}\right)
\qquad\quad
B_1=\left( \begin{array}{cc}  0&3\\ 1&0\\ 2&1 \\ 3&2\end{array}\right)
\end{equation}
\begin{equation}\label{bstateb}
B_2=\left( \begin{array}{cc}  0&2\\ 1&3\\ 2&0 \\ 3&1\end{array}\right)
\qquad\quad
B_3=\left( \begin{array}{cc}  0&1\\ 1&2\\ 2&3 \\ 3&0\end{array}\right)
\end{equation}
These matrices satisfy the rules in postulates 2 and 5.  This
measurement on two particles will play the same role the Bell
measurement plays in quantum teleportation.  Furthermore, it can be used
to prepare the analogue of the four Bell states. Imagine the two
particles are initially in some unknown state.  Now a $B$ measurement is
performed.  If the outcome is 0 corresponding to $B_0$ then the state
will be given by one of the rows in the $B_0$ matrix.  Looking at this
matrix we see that the two particles will become
correlated and their state will be $(y,y)$ where $y$ is unknown.
Similar remarks apply to the other outcomes.  For the outcome $r$
corresponding to $B_r$ the state will be $(y, y-r \mod 4)$.  Any of
these states can be changed into any other by applying the appropriate
rotation to one particle using the operation defined in (\ref{uk}).

\section{A no-cloning theorem}

As stressed earlier, a no-cloning theorem is essential to give meaning
to teleportation.  We will now prove that within our toy local theory
it is not possible to produce clones of particles in unknown states.
We will take an operational approach to cloning.  Thus, imagine that Peter
prepares a particle in some state by making a measurement on it.  As a
result of his measurement he might know that the state is something like
[50\% of 0 and 50\% of 1].  He now gives this particle to Alice. Alice
does not know what measurement Peter used to prepare the particles.  Alice
is challenged to give two particles back to Peter which are, so far
as Peter can establish, in the same state as the original.  One of these
two particles is to be regarded as the original and the other as the
clone.  Peter will then repeat the same measurement he used to prepare
the particles to test how well Alice has done.  If he gets the same
outcome as he did originally but now for both particles then Alice has
passed the test.  This test is repeated a large number of times.  Alice must
pass each time to convince Peter that she can produce clones.  We will
now prove that this is impossible -- Alice must fail this test sometimes.

Assume Peter prepares the state by measuring $P$ where
\[  P_0=\left( \begin{array}{c} 0 \\ 1 \end{array} \right)
    \qquad
    P_1=\left( \begin{array}{c} 2 \\ 3 \end{array} \right)  \]
He could, of course, have chosen any other measurement.  Assume further
that he gets outcome 0.  This means that the state is, from his point of
view, [50\% of 0 and 50\% of 1].  Lets assume that actually the state of
the particle is 1 (though Peter will not know this).  Peter passes the
particle on to Alice.  Alice receives a particle in state 1 (though she
does not know its state is 1) but no information about how it was
prepared.  Imagine now that on another occasion that Peter prepares
the state by measuring $P'$ where
\[  P'_0=\left( \begin{array}{c} 0 \\ 3 \end{array} \right)
    \qquad
    P'_1=\left( \begin{array}{c} 1 \\ 2 \end{array} \right)  \]
and he gets outcome 1.  This means that the prepared state, from Peter's
point of view, is [50\% of 1 and 50\% of 2].  We can assume that the
state of the particle is actually 1 (again Peter will not know this).
Peter passes the
particle on to Alice.  Again, Alice receives a particle in state 1
(though she does not know this) and no information about how it was
prepared.  From Alice's point of view these two cases are identical.
The only way she can hope to pass the test in all such cases is to
prepare two particles in the state 1.

To do this let us imagine that
she has $N-1$ particles.  When she receives the particle from Peter she
has $N$ particles.  She will perform some operations on the particle and
then give two particles to Peter who will perform his test.
There are only two types of operation she can
perform on the particles.  She can perform measurements on all the
particles as described in
postulates 2--5 and she can perform manipulations on individual
particles as described by postulate 7.  It is clear that the
manipulations described in postulate 7 cannot help since
they only act on individual particles and effectively do nothing more
than relabel the states.  Hence, assume Alice makes a measurement
$C$.  This measurement will have a certain number of outcomes.  In the
particular case we are discussing we want to produce two particles in
the state 1.  Hence, for at least one outcome, $r_1$ say, the matrix $C_{r_1}$
must have at least two 1's in at least one row (say the first row)
and the positions of
these 1's in this row must be known by Alice (since she must know which
particles to give to Peter).  Assume that the 1's are in the first two
columns (so that, in the case of obtaining this outcome, Alice will give
Peter particles 1 and 2).  Then the matrix will have the form
\begin{equation}
C_{r_1}= \left( \begin{array}{ccc} 1&1&\cdots \\
                                   u&v&\cdots \\
                                   \vdots&\vdots&\ddots \end{array}
                                   \right)
\end{equation}
In the first column at least 25\% of the entries must be different to 1
by postulate 5.  Assume that $u$ is one such entry ($u\not=1$).  If the
outcome is $r_1$ then after the measurement the state will randomly
selected from one of the rows in $A_{r_1}$.  However, at least 25\% of
these are of the form $(u,v,\dots)$ where $u\not=1$. Hence, when Alice
gives the first two particle to Peter, they will sometimes fail the
test because they will not always be in the state $(1,1)$.
This proves that Alice cannot successfully clone 100\% of the
time.

\section{Teleportation in the toy local theory}

To perform teleportation two particles, 2 and 3, are prepared in a Bell-like
state as described in section (\ref{bell}). This state is of the form
$(y,y)$ where $y$ is unknown.  The two particles separate from the
place of preparation.  Particle 2 goes towards Alice, and particle 3 goes
towards Bob. Alice is given a particle in an unknown state $x$.  She
takes this particle which we will call particle 1 together with particle
2 and performs the Bell-like measurement $B$ defined in
(\ref{bstatea},\ref{bstateb}).
She will get some outcome $r$ for this measurement. If the outcome is
$r=0$ we see from the matrix $B_0$ in equation (\ref{bstatea}) that $x$
for particle 1 is equal to
$y$ for particle 2.  In general, if she gets outcome $r$ then we see
from the matrices $B_r$ that $y=x-r\mod4$.
Particle 3 is in the state $y$.  Alice sends the information
$r$ to Bob and he can perform the
manipulation $U_r$ defined in (\ref{uk}) on particle 3.  After this its
state will be $x$.  This means that, without knowing the state of
particle 1, we have successfully transferred its state onto particle 3.
This is teleportation.  The state of particle 1 is completey randomised
after the Bell-like measurement and so the analogue with quantum
teleportation is very strong.

\section{Discussion}

We have shown that it is possible to have a local theory which has a
no-cloning theorem and supports teleportation.  The particular model
here provides some insight into what is essential for teleportation.
What happens in this model is the following.  The state of particle 1
is probed at a `microscopic'  level by particle 2 in a way that no
`macroscopic'  measuring apparatus could probe it.
It is this that
makes it possible to extract full information about the state of
particle 1 without running into the no-cloning theorem.  This information
is divided between the measurement result $r$ which Alice and Bob can know
and the state of particle 3 which they cannot know and which was
originally classically correlated with the `probe'   particle 2.
The original state can then be reconstructed in an analogous way to the
way it is reconstructed in quantum teleportation.
Nonlocality plays no role in this process.

Given this model, we must not assume that nonlocality is playing an
essencial role in quantum teleportation, though it may be the case that
it is.  There are two reasons for believing that nonlocality may play an
important role in quantum teleportation:  (1) As has been pointed out by
Bennett \cite{bennett}, the amount of information needed to specify a general
qubit is much greater than the two bits of information which is
classically communicated during quantum teleportation.  One might
speculate that, when a qubit is teleported, the extra information is
being carried by the nonlocal
properties of the entangled state.  On the other hand, it is not
possible to extract more classical information from a qubit than the two
classical bits communicated during teleportation and so there must
remain questions about the reality of the quantum information apparently
transmitted during teleportation.  (2) If one qubit of an
entangled pair is teleported it is possible to obtain a violation of
Bell's inequalities between the second qubit of this pair and the
teleported qubit. Hence, the teleportation machine must be able to
convey what ever quality is necessary for this nonlocal correlation.
However, we cannot necessarily assert on the basis of this fact that
nonlocality plays a role in quantum teleportation. It is possible that
the extra information which establishes the nonlocal correlations is
only transmitted in the process of measuring the quantities in Bell's
inequalities, and is not transmitted in the teleportation process.

Even if it is eventually concluded that quantum teleportation is a
nonlocal process, examples like that of Popescu \cite{popescu} suggest
that there may be a regime in which imperfect teleportation is happening,
but no nonlocality is in involved.  In such a regime teleportation would
still  circumvent the no-cloning theorem by probing the system to be
teleported at a microscopic level in much the same way as the toy local
theory considered in this paper.  The Werner states considered by
Popescu have since been shown to be nonlocal if more complicated
measurements are considered than those originally considered by Werner
(Popescu \cite{popescub}).  If two-particle
quantum states were found which are local and yet useful in
teleportation then we could conclude that the type of process discussed
in this paper also happens in quantum theory.  States having bound
entanglement \cite{horos} are obvious candidates to consider.
However, while it seems likely, it has not yet been shown
that such states will always satisfy Bell type inequalities.
Furthermore, it is not known whether such states could be useful for
teleportation although results derived so far are negative \cite{linpop}.

Whatever conclusions may eventually be drawn about nonlocality in
quantum teleportation, it has been established in this paper that
teleportation in theories which have a no-cloning theorem can be an
entirely local phenomenon.


\begin{thebibliography}{99}

\bibitem{bella} J. S. Bell, Physics {\bf 1}, 195 (1964)
\bibitem{teleport} C. H. Bennett, G. Brassard, C. Crepeau, R. Jozsa, A.
Peres, and W. K. Wooters, Phys. Rev. Lett {\bf 70}, 1895 (1993)
\bibitem{popescu} S. Popescu, Phys. Rev. Lett. {\bf 74}, 2619 (1995).
\bibitem{braunstein}. S. L. Braunstein and  H. J. Kimble, Phys. Rev.
Lett. {\bf 80}, 869 (1998).
\bibitem{bellb} J. S. Bell, chap 21 in {\it Speakables and unspeakable
in quantum mechanics}, (CUP, 1987).
\bibitem{cohen}  O. Cohen, Phys. Rev. A {\bf 56}, 3484 (1997).
\bibitem{expmts} D. Boschi, S. Branca, F. De Martini, L. Hardy, and S.
Popescu, Phys. Rev. Lett. {\bf 80}, 1121 (1998), D. Boumeester, J. W.
Pan, K. Mattle, M. Eibl, H. Weinfurter, and A. Zeilinger, Nature {\bf
390}, 575(1997),
A. Furusawa, J. L. Sorensen, S. L. Braunstein, C. A. Fuchs, H. J.
Kimble, and  C. A. Fuchs, Science {\bf 282}, 706 (1998).
\bibitem{ramon}R. Risco Delgado, The meaning of teleportation in terms
of quantum zeropoint field, preprint.
\bibitem{bennett} C. H. Bennett, private communication.
\bibitem{popescub} S. Popescu, Phys. Rev. Lett. {\bf 74}, 2619 (1995).
\bibitem{horos} P. Horodecki, Phys. Lett. A {\bf 232}, 333 (1997); M.
Horodecki, P. Horodecki, and R. Horodecki, ``Mixed-state entanglement and
distillation: is there `` bound'' entanglement in nature'',
LANL e-print quant-ph/9801069.
\bibitem{linpop} N. Lindon and S. Popescu, ``Bound entanglement and
teleportation'' LANL e-print quant-ph/9807069.

\end{thebibliography}
\end{document}